\documentstyle[prl,aps,preprint]{revtex}

\begin{document}
\draft

\title{
Randomness at the Edge:\\
Theory of Quantum Hall transport at filling $\nu=2/3$ }

\author{C.L. Kane}

\address{Department of Physics, University of Pennsylvania\\
Philadelphia, Pennsylvania 19104
}
\author{M.P.A. Fisher and J. Polchinski}
\address{Institute for Theoretical Physics, University of California\\
Santa Barbara, CA 93106
}

\date{\today}
\maketitle

{\tightenlines
\begin{abstract}
Current Luttinger liquid edge state theories for filling $\nu=2/3$ predict a
non-universal Hall conductance, in disagreement with experiment.
Upon inclusion of random edge tunnelling we find a phase transition
into a new disordered-dominated edge phase.  An exact solution of the
random model in this phase gives a quantized Hall conductance of 2/3 and
a neutral mode propagating upstream.
The presence of the neutral mode changes the predicted temperature
dependence for tunnelling through a point contact from $T^{2/\nu -2}$ to
$T^2$.

\end{abstract}
}

\narrowtext

The most striking aspect of the quantum Hall effect is the remarkable
quantization of the Hall conductance measured in transport experiments
\cite{Klitzing,Tsui}.
For the integer
quantum Hall effect, the edge state picture formulated by Halperin
\cite{Halperin}
has been incorporated into the framework of Landauer-Buttiker
transport theory\cite{Landauer,Buttiker},
and provides an extremely
simple explanation of this quantization.
An edge state explanation for the quantized conductance in the
fractional quantized Hall effect (FQHE), though, is more subtle.  An important
advance was made by Wen\cite{Wen}, who argued that
edge excitations in the FQHE can be
described at low energies by a chiral Luttinger liquid model.  For the
Laughlin states, such
as $\nu=1/3$, where the edge state has only one branch, this
approach gives a simple explanation of the Hall quantization, and moreover has
enjoyed success in describing
tunneling between edge states\cite{Kane,Webb}.
However, for hierarchical quantum
Hall states, such as $\nu=2/3$,
the edge state structure is much more complicated and the whole
approach more problematic.  Indeed, Wen\cite{Wen} and MacDonald\cite{MacDonald}
have argued that a
$\nu=2/3$ edge
consists of two edge branches which move in $\it{opposite}$ directions.
However, a recent time domain experiment\cite{tdomain}
has shown only $\it{one}$ propagating mode at the $\nu=2/3$ edge.
Moreover, including electron interactions between
the two branches in this model gives a Hall conductance which
is $\it{not}$ quantized and {\it non-universal}, depending on the
strength of the interactions. Is something seriously amiss with
the edge state approach?

In this letter we re-consider the $\nu=2/3$ edge and argue that the
problem
lies not with the edge state approach, but with the assumption of an ideal
edge.
We show that in the presence of random edge scattering,
there is an edge phase transition at a critical interaction strength from
the state in
which the zero temperature conductance is
non-universal, to a new disorder-dominate edge phase.  By exploiting a hidden
SU(2) symmetry, we obtain an exact solution
of the model in the disorder-dominated phase which reveals only a single
propagating charge mode.  This mode gives a quantized
Hall conductance of 2/3.   Thus, random edge scattering is apparently
necessary to explain the Hall quantization at filling
$\nu=2/3$!  The exact solution reveals, moreover, the presence of
a $\it{neutral}$ edge mode, which propagates in the direction opposite to the
charge mode.  The neutral mode, which does not contribute to the Hall
conductance,
would not be detectable in the capacitive time domain
experiments of Ref. \onlinecite{tdomain}.
However, we show that the neutral mode $\it{modifies}$ the temperature
dependence of the conductance through
a point contact in a $\nu=2/3$ fluid, changing the behavior from
$G(T)\sim T^{2/\nu -2}$, to $G(T)\sim T^2$.  We suggest a
variant of the time-domain
experiment of Ref. \onlinecite{tdomain}
which should allow a direct observation of the neutral
mode.

McDonald and Wen \cite{Wen,MacDonald} have argued that the
$\nu=2/3$ edge consists of two modes: A forward propagating
edge mode, similar to a $\nu=1$ edge, and a backward propagating
mode, similar to a $\nu=1/3$ edge.  This picture is supported
by microscopic calculations for a sharp edge which show a non-monotonic
electron density,
increasing from  $\nu=2/3$ to $\nu=1$,
before falling to zero.  Following Wen, the appropriate (Euclidian) action is:
\begin{equation}
S_0 =  \int dx d\tau {1\over 4\pi} \left[
\partial_x \phi_1 ( i\partial_\tau + v_1 \partial_x) \phi_1 +
3 \partial_x \phi_2 ( -i\partial_\tau + v_2 \partial_x) \phi_2
- 2v_{12} \partial_x \phi_1 \partial_x \phi_2 \right]  .
\end{equation}
When $v_{12}=0$, this describes two independent chiral Luttinger liquids,
equivalent to the edge states of the $\nu=1$ and $\nu=1/3$ quantum
Hall fluids.
The 1d electron charge density in each mode is given by
$\pm \partial_x \phi_{1,2}$.
The velocities $v_1$ and $v_2$
reflect the interactions within each channel.
We suppose the interactions are short ranged, screened by a ground plane. In
general, there will
also be a repulsive interaction $v_{12}>0$ between the two channels.
This mixes the channels together, giving eigenmodes that are linear
combinations of $\phi_1$ and $\phi_2$.
Stability requires that $v_{12}$ is
not too large, namely $v_{12}^2<3v_1v_2$.

The two-terminal conductance of a $\nu=2/3$ Hall bar with top and
bottom edges described
by (1) may be computed from the Kubo formula,
$G = <I(\omega_n)I(-\omega_n)>/\hbar|\omega_n|$.  Here the current
$I$ is a sum of the currents on the top and bottom edges at $x=0$,
each given by $ie\partial_\tau (\phi_1 - \phi_2)/2\pi$.
We find
\begin{equation}
G = {2\over 3} \Delta {e^2\over h}   ;\qquad
\Delta = (2-\sqrt{3}c)/\sqrt{1-c^2}
\end{equation}
with $c=(2v_{12}/\sqrt{3})(v_1+v_2)^{-1}$.
The stability requirement implies that $|c|<1$.
Notice that G is non-universal, depending on the
interaction strengths\cite{Coulomb}.
A similar calculation shows that the four-terminal Hall conductance is also
non-universal,
given by $G_H = (1/3)(1+\Delta^2) e^2/h$.
There is one special choice of interactions,
$c=\sqrt{3}/2$, for which both $G$ and $G_H$ are
minimal and given by $G_{min}=2/3$.  But experiments need no
such fine tuning!

The absence of a quantized conductance can be traced to a lack of equilibration
between the right and left moving
edge channels.  To see this consider the case of de-coupled
channels, $v_{12}=0$, for which $G = 4/3$, rather than $2/3$.  The value
$4/3$ also follows from a simple
Landauer type argument, in which current from the source injected
into the $\phi_1$ mode on, say, the top edge of a Hall bar is added to
another current
injected from the source into the $\phi_2$ mode on the $\it{bottom}$ edge.
Clearly, this requires that the
two opposite moving modes on a given edge do $\it{not}$ equilibrate
with one another.  To allow for possible equilibration, the interaction
$v_{12}$ is also insufficient.  Rather, it is
essential to allow the electrons to tunnel between the two edge modes.
However, a (spatially) constant inter-mode tunnelling term is ineffective at
backscattering, since the two modes will generally have different momenta.
(The momentum difference, which is gauge invariant, is proportional to
the magnetic flux penetrating between the two edge modes.)
But if the inter-mode tunnelling is (spatially) $\it{random}$,
momentum along the edge is not conserved, and backscattering
(and hence equilibration) is possible.

Consider then random electron tunnelling between the two modes $\phi_1$ and
$\phi_2$.
The operator $\exp(i\phi_1)$ adds an electron to the mode $\phi_1$,
whereas $\exp(i\phi_2)$ adds a 1/3 charge Laughlin quasiparticle to the mode
$\phi_2$.  Since inter-edge tunnelling conserves charge, the simplest
(and most relevant) term is $\exp(i\phi_1 - 3i \phi_2)$, which
hops three quasiparticles, with total charge $e$, from mode 2 to mode 1.
The corresponding term to add to the action (1) is thus:
\begin{equation}
S_1= \int dx d\tau \xi(x) \exp (i\phi_1 - 3i\phi_2) + c.c.
\end{equation}
In general $\xi$ is complex.  Indeed, for (spatially) uniform tunnelling
one would have $\xi = t \exp(ik_{12}x)$ with $k_{12}$ the momentum
difference between the two modes.  But since this oscillates rapidly as
x-varies it is unimportant at long lengthscales.  For a
realistic edge with impurities, $\xi(x)$
 will be random
and uncorrelated at large separations.  For simplicity we assume
that $\xi(x)$ is a gaussian random
variable satisfying $\overline{\xi^*(x)\xi(x')} = W \delta(x-x')$.

For small randomness $W$ the effect of the tunneling term can be analyzed by
evaluating the pair correlation function
of the tunnelling operator $O=\exp(i\phi_1-3i\phi_2)$ using the quadratic
action (1).  We find:
 \begin{equation}
\langle O(x,\tau) O(0,0) \rangle_0 \sim
{{1}\over{(v_+\tau + ix)^{\Delta+1}}} {{1}\over { (v_-\tau -ix)^{\Delta-1}}},
\end{equation}
with $\Delta$ given in (2).  The total scaling dimension of the operator,
a sum of right and left moving dimensions, is equal to $\Delta$.
The difference between the dimensions is 2, consistent with boson statistics
of the operator $O$.

Under a renormalization group (RG) transformation which leaves the
quadratic action (1) invariant, the dimension $\Delta$
determines the relevancy of $O$.  Since the perturbation
(3) is spatially random\cite{Giamarchi}, the leading RG flow equation for $W$
is:
\begin{equation}
{dW\over{d\ell}} = (3-2\Delta)W  .
\end{equation}
For $\Delta>3/2$, which corresponds to
small interaction $v_{12}$, small randomness is irrelevant.
In this case at $T=0$, the
conductance $G=2\Delta/3$ is non-universal\cite{equil},
with $G>1$.  At the transition $G^*=1$.  For $\Delta<3/2$, weak
random tunnelling grows and drives the edge into
a disorder dominated phase.

Can this disorder-dominated phase correctly describe a real $\nu=2/3$
edge with quantized Hall conductance?  Clearly perturbation theory in $W$
is useless, and a non-perturbative approach is
necessary.  Since our model is a chiral generalization
of an interacting 1d localization problem this might seem hopeless.  However,
we now show that the strong coupling
phase has a hidden SU(2) symmetry, which allows for a complete solution.  To
this end introduce first new fields:

\begin{equation} \phi_{\rho} = \sqrt{3/2} (\phi_1 - \phi_2) ;\qquad
 \phi_{\sigma} = \sqrt{1/2} (\phi_1 - 3\phi_2)   .
\end{equation}
Here $\phi_{\rho}$ is a "charge" mode
and $\phi_{\sigma}$ is a "neutral" mode.
The charge current along an edge is $i\partial_\tau \phi_\rho/2\pi$.
The total action $S=S_0+S_1$ can be re-expressed as
$S=S_{\rho}+S_{\sigma}+S_{pert}$
with charge and neutral pieces

\begin{equation}
S_{\rho} =  {1\over 4\pi} \int_{x,\tau}
\partial_x \phi_{\rho} ( i\partial_\tau + v_{\rho} \partial_x) \phi_{\rho}
\end{equation}
\begin{equation}
S_{\sigma} = {1\over 4\pi} \int_{x,\tau}
\partial_x \phi_{\sigma} ( -i\partial_\tau + v_{\sigma}
\partial_x) \phi_{\sigma} + \int_{x,\tau} [\xi(x)e^{i\sqrt{2}
\phi_{\sigma}} + c.c.]
\end{equation}
coupled together via
\begin{equation}
S_{pert} = -v {1\over 4\pi} \int_{x,\tau} \partial_x \phi_{\rho}
\partial_x \phi_{\sigma}   .
\end{equation}

The velocities, $v_\rho$, $v_\sigma$ and $v$ depend on the original
velocities in (1) in a complicated way (which we don't display), but for
$\Delta-1$ small
are simply related to one another:  $v = (v_{\rho}+v_{\sigma})
\sqrt{(\Delta-1)/2}$.  Thus at
$\Delta=1$, the interaction $v$ vanishes, and the charge and neutral
modes de-couple.  As we now show, this de-coupled problem can be
solved exactly for arbitrary randomness $W$, giving
a fixed line at $\Delta=1$ in the $\Delta-W$ plane (see
Figure).  The coupling term, $S_{pert}$, is then shown to be
an irrelevant perturbation, so that the RG flows are towards the de-coupled
fixed line.

When $v=0$ the two operators $\cos(\sqrt{2} \phi_{\sigma})$ and
$\sin(\sqrt{2} \phi_{\sigma})$ in $S_\sigma$
both have scaling dimension $\Delta=1$, as does
$\partial_x \phi_{\sigma}$.  These three operators together satisfy an SU(2)
algebra, known as a level-one $SU(2)$ current algebra.
The quadratic part of the action $S_{\sigma}$ respects this
SU(2) symmetry, whereas the non-linear terms are random SU(2) symmetry breaking
fields.  This SU(2) symmetry allows us to write
down a model in terms of a two component fermion $\psi$, which has
identical low-energy physics to $S_{\sigma}$.
We first introduce an $\it{extra}$ bosonic field $\chi$ with free action
$S_\chi$ the same as that of $\phi_\sigma$ in the first term of equation (8).
The action $S_\sigma+S_\chi$ is then equivalent to the bosonized
representation of a model of spin 1/2 chiral fermions,
\begin{equation}
S_{\psi} = \int_{x,\tau} \psi^\dagger ( \partial_\tau +iv_\sigma \partial_x )
\psi  + \psi^\dagger (\xi \sigma^+ + \xi^\ast \sigma^- ) \psi  ,
\end{equation}
provided we identify
$\psi_1 = \exp[i(\chi + \phi_\sigma)/\sqrt{2}]$
and
$\psi_2 = \exp[i(\chi-\phi_\sigma)/\sqrt{2}]$.
Here $\sigma^\pm = \sigma_x \pm i \sigma_y$ with Pauli matrices
$\sigma_\mu$.  The operator corresponding to $\partial_x \phi_\sigma$
is $\psi^\dagger \sigma_z \psi$.
Note that $\chi$ does not enter physical observables, but
allows a convenient representation of the $SU(2)$ symmetry.

The random terms can now be completely eliminated from the action
(11) by performing
a unitary SU(2) gauge transformation, $\tilde{\psi}(x) = U(x) \psi(x)$,
with
\begin{equation}
U(x) = T_x \exp[-i \int_{-\infty}^x d x^\prime M(x^\prime )] ,
\end{equation}
where the 2x2 matrix $M(x)=[\xi(x) \sigma^+ + \xi^*(x) \sigma^-]/v_\sigma$ and
$T_x$ is an x-ordering
operator.  In terms of the new field the action is that for free (chiral)
fermions:
\begin{equation}
S_{\tilde{\psi}} = \int_{x,\tau} \tilde{\psi}^\dagger (
\partial_\tau +iv_\sigma \partial_x )
\tilde{\psi}  .
\end{equation}
Taken together, (7) and (12) represent a complete solution of the de-coupled
line, $\Delta=1$.
Since both actions are quadratic, the line is in fact a
{\it fixed-line}.
Notice that $S_{\tilde{\psi}}$ describes a $\it{propagating}$ neutral mode.
The model does $\it{not}$
exhibit localization.  The original neutral mode, $\phi_\sigma$ in (8),
is not conserved with randomness and does not propagate.

Consider now $S_{pert}$ which couples together the neutral and charge sectors.
In terms of $\tilde{\psi}$ this becomes:
\begin{equation}
S_{pert} = -v {{1}\over{4\pi}} \int_{x,\tau}
\partial_x \phi_\rho \tilde{\psi}^\dagger U(x) \sigma_z U^\dagger(x)
\tilde{\psi} .
\end{equation}
The relevancy of this term on the fixed line described by
$S_\rho + S_{\tilde\psi}$ can be obtained from the scaling dimension of the
operator, which we denote $\delta_v$.  Using (7) and (12) one readily
obtains $\delta_v=2$.  Since the operator
has a random x-dependent coefficient, $U\sigma_z U^\dagger$,
which is uncorrelated on spatial scales large compared to $v_\sigma^2/W$,
the linear RG flow equation for
$v^2$ is $\partial (v^2)/ \partial\ell = (3-2\delta_v) v^2$.  Thus
$v$ is irrelevant.  Since $v^2 \sim \Delta -1$, the RG flows are
towards the fixed line at $\Delta=1$
 (see Figure).

While we cannot ascertain the precise domain of attraction of the $\Delta =1$
fixed
line, some information can be obtained by
performing a perturbative RG near the phase transition, when
$W$ and $\Delta-3/2$ are small.
The renormalization of $W$ is described by (5).
In terms of the right and left moving velocities, $v_\pm$ in (4),
the remaining RG equations linear in $W$ are,
\begin{eqnarray}
{d\Delta\over{d\ell}} & = & -8\pi {\sqrt{v_+ v_-^{-3}}\over{v_+ + v_-}}
(\Delta^2 - 1)W   ,\\
{dv_\pm\over{d\ell}} & = & -4\pi {v_\pm^2 \over \sqrt{v_+ v_-^5}}
(\Delta \mp 1) W   .
\end{eqnarray}
Note that the transition has a Kosterlitz-Thouless (KT) type
form, with $\Delta$ being driven downward for $W\ne 0$ (see Figure).  In view
of this it is most plausible that
all of the flows above the KT-sepatrix eventually make their way to the
$\Delta=1$ fixed line.

Since the charge and neutral sectors de-couple on the $\Delta=1$ fixed line,
only the charge mode (7) contributes to the conductance giving the
quantized value, $G=2/3$.
The corrections to quantization due to the irrelevant variable $v$ may be
computed perturbatively using the exact solution above.
It is noteworthy that in the
D.C. limit, the correction vanishes, so that there is {\it no} power law
finite temperature correction.  At finite frequency the correction is
$\Delta G(\omega) \sim v^2 \omega^2/ (\omega + i/\tau)^2$.
We interpret $\tau\sim 1/W$ as an equilibration time
for the two edge modes\cite{equil}.

Despite carrying no charge, the upstream propagating neutral
mode can be detected in at least two ways.  The first, which is less direct,
involves
tunnelling through a constricted point contact in a
$\nu=2/3$ Hall
fluid.  For filling $\nu^{-1}$ an odd integer, where the edge mode
has only one branch, it was shown that
the conductance through a point contact vanishes with temperature as $G(T) \sim
T^{2/\nu -2}$.
For $\nu=2/3$ one might therefore expect a power law, $G(T)\sim T$.
However, the presence of the neutral mode at the $\nu=2/3$ edge
increases this power by one, giving the prediction $G(T) \sim T^2$,
as we now show.

The most general edge tunnelling operator at a $\nu=2/3$
edge is $O_{n_1,n_2} = e^{i(n_1\phi_1 +
n_2 \phi_2)}$, with integer $n_1$ and $n_2$.  This operator
creates an edge excitation with charge, $Q(n_1,n_2)=e(n_1+n_2/3)$.
For tunnelling through a point contact the important quantity is
the "local" dimension, $\delta$, of the operator,
$\delta(n_1,n_2)$, defined via
$<O(x,\tau)O(x,0)> \sim \tau^{-2\delta}$.
Being at one spatial point, this average is independent of the
random SU(2) gauge transformation (11), and $\delta$ can be readily
evaluated using the free action (7) and (12).  We find,
\begin{equation}
 \delta(n_1,n_2) = {3\over4} (n_1+{n_2\over3})^2 + {1\over4}(n_1+n_2)^2
\end{equation}
where the first and second contribution are from the charge and neutral
sectors respectively.  Let $t$ be the amplitude for tunnelling
an electron
through the point contact.  The RG equation for
$t$ is $\partial t / \partial \ell = (1-2\delta)t$, with
$\delta$ the dimension of the charge $e$ operator which minimizes (16), namely
$\delta=\delta(1,0)=1$.  This gives an effective temperature dependent
amplitude, $t_{\it eff}(T) \sim t T^{2\delta-1}$, and
a conductance which varies as
$G(T) \sim t_{\it eff}^2 \sim t^2 T^2$.  If the neutral mode were absent,
one would have $\delta=3/4$, and hence $G(T) \sim T^1$.
Thus, a measured $T^2$ temperature dependence would give (indirect)
evidence of the neutral mode.

Time domain transport through a large quantum dot
at filling $\nu=2/3$, might enable a much more direct measurement of
the neutral mode.  Imagine two leads
coupled via tunnel junctions to
opposite sides of a dot.  A short current pulse incident in one lead,
upon tunnelling into the dot, would
excite both the charge and neutral edge modes. These excitations, after
propagating along the edge of the dot in
opposite directions and at different velocities, would, upon arrival at the far
tunnel junction, excite
$\it{two}$ current pulses into the outgoing
lead.  By tailoring the placement of
the leads, a measurement
of the direction of propagation of the neutral should also be possible.

In conclusion, we have solved exactly a model of a
disordered $\nu=2/3$ edge.  This solution enables an analysis of
resonant tunnelling and non-equilibrium transport.  This analysis, and a
generalization
to other composite FQHE edges, such as $\nu=4/5$, will be reported elsewhere.

M.P.A.F. is grateful to the Aspen Center of Physics where
this work was initiated, and to Alan MacDonald for numerous
illuminating converations.
This work has been supported by the NSF under grants PHY-90-09850 and
PHY-91-57463.

\begin{figure}
\caption{Renormalization group flow diagram as a function of disorder
strength $W$ and the scaling dimension $\Delta$ of the tunneling operator.
For $\Delta<3/2$ all flows end up at the exactly soluble fixed line
$\Delta=1$.}
\label{fig1}
\end{figure}


\begin{references}
\bibitem{equil}
The conductance for $\Delta>3/2$ at $T\ne0$ is tricky.  Further analysis
reveals that the irrelevancy of disorder $W$ leads to an
equilibration length which diverges for $T \rightarrow 0$ as
$\ell_{in}(T) \sim [WT^{2(\Delta -1)}]^{-1}$.  For a system of length $L$,
the conductance will be non-universal at low $T$, with $G=2\Delta/3$,
but might crossover to the quantized value $G=2/3$ when
$\ell_{in}(T) << L$.
\end{references}
\end{document}